\documentclass[letterpaper,prl,twocolumn,preprintnumbers]{revtex4}
\usepackage{graphicx,bm}
\begin{document}
\newcommand{\mybm}[1]{\mbox{\boldmath$#1$}}
\newcommand{\mysw}[1]{\scriptscriptstyle #1}

\preprint{\sf Version 2 (\today)}
\title{Andreev Edge State on Semi-Infinite Triangular Lattice: Detecting the Pairing Symmetry in Na$_{0.35}$CoO$_{2} \cdot y$H$_{2}$O}
\author{T. Pereg-Barnea$^{1,2}$ and Hsiu-Hau Lin$^{1,3*}$}
\affiliation{$^{1}$Kavli Institute for Theoretical Physics, University of California, Santa Barbara, CA 93106\\
$^{2}$Department of Physics and Astronomy, University of British Columbia, Vancouver, B.C. V6T 1Z1, Canada\\
$^{3}$Physics Division, National Center for Theoretical
Sciences, Hsinchu 300, Taiwan}
\altaffiliation[]{On leave from Department of Physics, National Tsing-Hua University, Hsinchu 300, Taiwan}
\date{\today}

\begin{abstract}
We study the Andreev edge state on the semi-infinite triangular lattice with different pairing symmetries and boundary topologies. 
We find a rich phase diagram of zero energy Andreev edge states that is a unique fingerprint of each of the possible pairing symmetries.
We propose to pin down the pairing symmetry in recently discovered Na$_{x}$CoO$_{2}$ material by the Fourier-transformed scanning tunneling 
spectroscopy for the edge state. A surprisingly rich phase diagram is found and explained by a general gauge argument and mapping to 1D tight-binding model. Extensions of this work are discussed at the end. 
\end{abstract}
\maketitle

Recently, Takada {\it et al.} discovered that Na$_{0.35}$CoO$_{2} \cdot y$H$_{2}$O undergoes a superconducting phase transition around 5 K\cite{Takada03}. 
Despite the low critical temperature, it triggered intense attentions from both the experimental\cite{Sakurai03,Lorentz03,Schaak03,Cao03,Jin03,Chou04} and theoretical\cite{Baskaran03,Koshibae03,Tanaka03,Wang04,Ikeda04,Tanaka04} sides of the scientific community due to its similar planar structure to the high $T_{c}$ materials. In contrast to the square lattice in cuprates, the Co atoms form a two-dimensional triangular lattice without orbital degeneracy.

Since the antiferromagnetic correlations are frustrated in the triangular lattice, Baskaran\cite{Baskaran03} explored the implications of resonant valence bond (RVB) physics in this material. Based on the RVB picture, theoretical investigations on the $t$-$J$ model\cite{Wang04} favor $d_{x^2-y^2}$+$id_{xy}$ symmetry. 
Taking a different perspective, Tanaka and Hu\cite{Tanaka03} proposed that the peculiar shape of Fermi surface stabilizes $p_x$+$ip_y$ spin-triplet pairing. 
On the other hand, starting from fluctuation-exchange approximation, the simple $f$-wave pairing is predicted\cite{Ikeda04}, which is also suggested by RPA calculations\cite{Tanaka04} and is strongly tied to the charge fluctuations in the system. Since various approximations are assumed in the above approaches, the pairing symmetry in Na$_{x}$CoO$_{2}$ remains controversial at the point of writing.

While it is important to determine the pairing symmetry from a microscopic approach, it is as important to understand the phenomenology associated with different pairing symmetries. In this Letter, we study the Andreev edge state (AES) localized at the boundary and propose to pin down the pairing symmetry by Fourier-transformed scanning tunneling spectroscopy (FT-STS)\cite{Hoffman02,McElroy03}. Due to growing evidence for the existence of nodes in the gap\cite{Ishida03,Fujimoto04}, we would concentrate on the $p$, $d$ and $f$ pairing symmetries here. Furthermore, since detailed calculations for different symmetries are rather similar, we only elaborate on the simplest $f$-wave pairing, which fits well with the hexagonal symmetry in the triangular lattice and does not break time reversal symmetry.

\begin{figure}
\centering
\includegraphics[width=7cm]{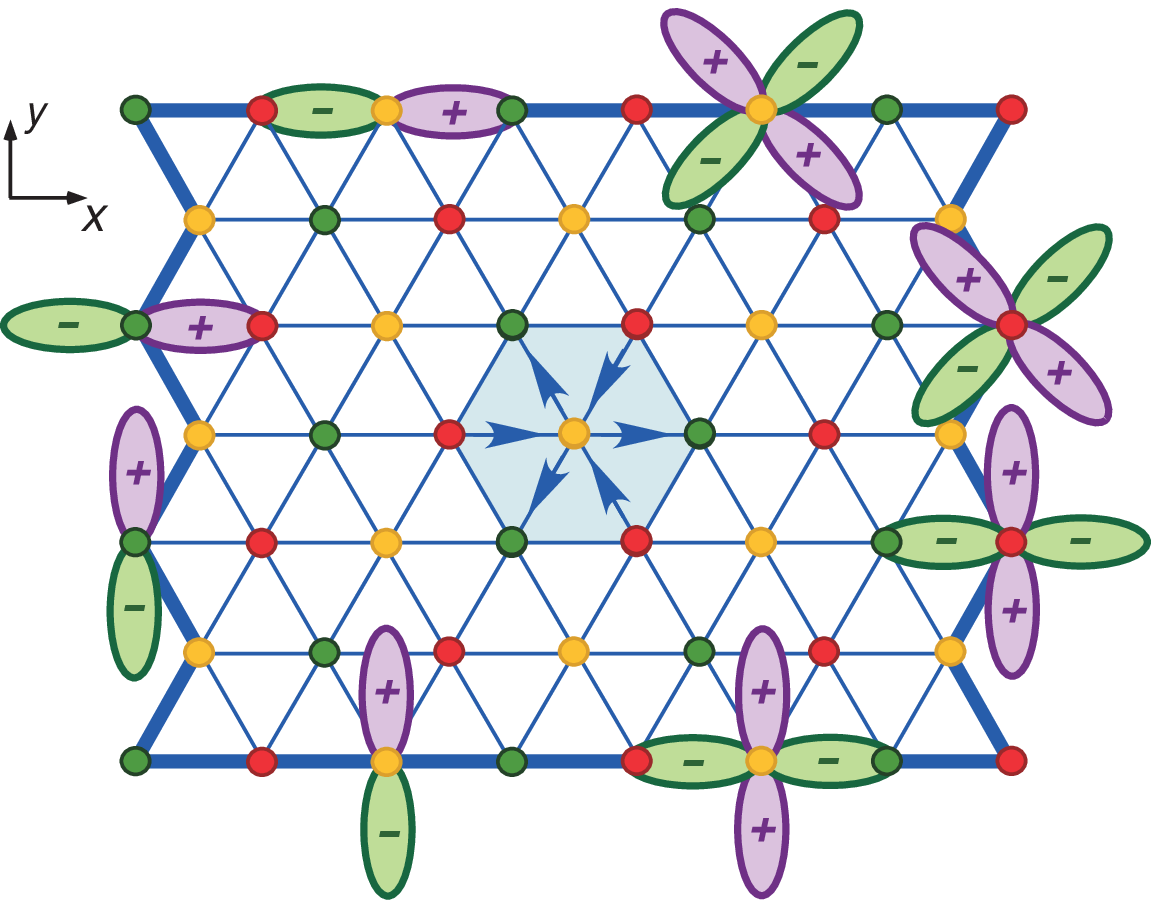}
\includegraphics[width=\columnwidth]{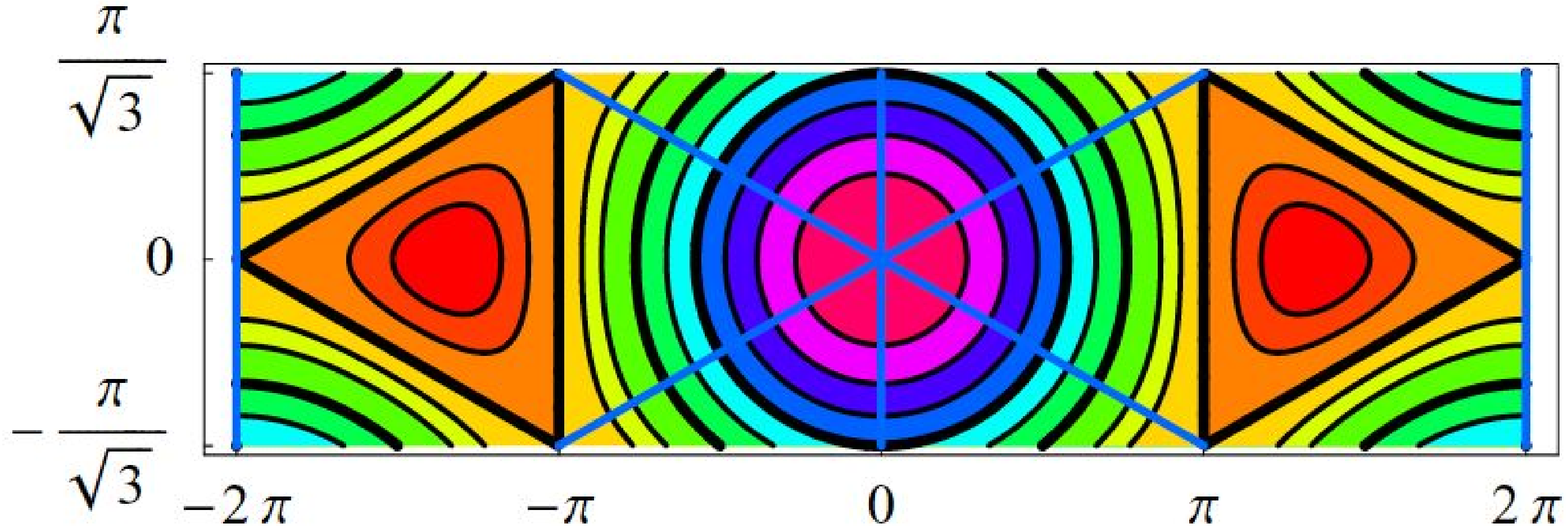} 
\caption{\label{fig:cuts} Gap function of $p_{x}$, $p_{y}$, $d_{xy}$ and $d_{x^{2}\mbox{-}y^{2}}$ symmetries at two different zigzag and flat edges for the triangular lattice. The $f$-wave pairing potential is shown in the middle shaded hexagon, with signs determined by either parallel ($\Delta>0$) or anti-parallel ($\Delta<0$) to the arrows. In the lower panel, the Fermi surface in the reconstructed Brillouin zone is shown. The thick contours are at different chemical potentials, $\mu/t=2$ (middle circle), $\mu/t=0$ (fragmented circle) and $\mu/t= -2$ (two triangles). The nodes are located at the intersections of the Fermi surface contours and the nodal lines.}
\end{figure}

Note that there are two natural boundary topologies on the triangular lattice -- zigzag and flat edges, as shown in Fig.~\ref{fig:cuts}. For $f$-wave symmetry, the Andreev edge state exists at zigzag edge for some momenta $k_{y}$ along the boundary while it is absent at the flat edge. The results for other pairing symmetries are summarized in Table \ref{PairingSymmetry}.

\begin{table}
\begin{tabular}{|c|c|c|}
\hline\hline 
zigzag edge &flat edge & pairing symmetry\\
\hline
Yes & No& $f$ or $p_{x}$ \\
\hline
No&Yes& $p_{y}$\\
\hline
Yes&Yes& $d_{xy}$\\
\hline
No&No&$d_{x^{2}\mbox{-}y^{2}}$ or $s$\\
\hline\hline
\end{tabular}
\caption{\label{PairingSymmetry} Existence of Andreev bound state at zigzag and flat edges and its implication for pairing symmetry.}
\end{table}

Recent breakthrough in FT-STS experiments allow further insight  into the edge state, beyond its detection only. In these experiments one probes the local density of states (LDOS) of a two dimensional sample surface on a large field of view that allows good resolution of the Fourier transformed data. In the case of AES, their exponential decay away from the boundary can be detected directly while their dependence upon the transverse momentum (along the edge where the system is translationally invariant) can be seen in Fourier space through scattering processes. In this Letter we show that different pairing symmetries give rise to edge states with different transverse momentum profiles. For instance, the structure of the LDOS for the $f$-wave pairing at zigzag edge gives rise to interference pattern of quasi-particle scattering which peaks at $Q=\pm \pi/\sqrt{3}a$ after the STS data is Fourier analyzed.

Another interesting aspect is the richness of the phase diagram at different momentum $k_{y}$, arising from the tripartite property of the triangular lattice. We develop a general mapping to 1D tight-binding model which combines the subtle interplay between the symmetry of the order parameter and the topology of the boundary together. The complicated phase diagram of AES can be understood as some intrinsic property of the hopping matrix in the effective 1D lattice. 
In addition, it is evident in the phase diagram that the different phases (whose nature will become clear shortly) are 
separated by the nodal extended states. This can be shown by Oshikawa's 
gauge argument\cite{Oshikawa00,Lin04}.

To obtain the phase diagram of AES, let us start with the Bogoliubov-de Gennes Hamiltonian for $f$-wave pairing with zigzag edge,
\begin{eqnarray}
H_{BdG} &=& t \sum_{\langle{\bf r, r'}\rangle, \alpha} c^{\dag}_{\alpha}({\bf r}) c^{}_{\alpha}({\bf r'}) - \mu \sum_{{\bf r}, \alpha} c^{\dag}_{\alpha}({\bf r}) c^{}_{\alpha}({\bf r})
\nonumber\\
&&\hspace{-15mm}+\sum_{\langle \bf r,r' \rangle} [\Delta^{*}({\bf r,r'}) c_{\uparrow}({\bf r}) c_{\downarrow}({\bf r'}) + \Delta({\bf r,r'}) c^{\dag}_{\uparrow}({\bf r}) c^{\dag}_{\downarrow}({\bf r'})].
\end{eqnarray}
Here only nearest-neighbor hopping and pairing are included. Since the triangular lattice is not particle-hole symmetric, the sign of the hopping amplitude is important. Recent experiments\cite{Singh00,Valla02} suggest that the maximum of the band occurs at the $\Gamma$ point, which implies $t>0$. The $f$-wave paring potential $\Delta({\bf r}, {\bf r'})= -\Delta({\bf r'}, {\bf r})=\pm \Delta$, with signs depending on the bond orientation 
as shown in Fig.~\ref{fig:cuts}. The anti-symmetric spatial dependence arises from spin triplet pairing.

The above Hamiltonian can be decomposed into a sum of 1D Hamiltonians by partially Fourier transforming in the edge direction ($y$). Note that the conventional hexagonal Brillouin zone must be folded into the particular rectangle (shown in Fig.~\ref{fig:cuts}) to keep the Fermi statistics\cite{Lin98}. Each 1D Hamiltonian can be conveniently written in the basis of bonding and antibonding of particle and hole creation operators,
\begin{eqnarray}
\Psi(k_{y}) = \frac{1}{\sqrt{2}}\left[ 
\begin{array}{c}
c_{\uparrow}(x,k_{y})+c^{\dag}_{\downarrow}(x,-k_{y})\\
-c_{\uparrow}(x,k_{y})+c^{\dag}_{\downarrow}(x,-k_{y})
\end{array}\right].
\end{eqnarray}
The Hamiltonian takes the simple (supersymmetric) Dirac form,
\begin{eqnarray}
H_{\mysw{BdG}}=\sum_{k_{y}} \Psi^{\dag}(k_{y}) \left( 
\begin{array}{cc}
0&A\\
A^{\dag}&0
\end{array}\right) \Psi(k_{y}),
\label{DiracH}
\end{eqnarray}
where the semi-infinite matrix $A$ is
\begin{eqnarray}
A = \left(
\begin{array}{cccccc}
-\mu&T_{1}&T_{2}&0&0&...\\
T_{\bar 1}&-\mu&T_{1}&T_{2}&0&...\\
T_{\bar 2}&T_{\bar 1}&-\mu&T_{1}&T_{2}&...\\
0&T_{\bar 2}&T_{\bar 1}&-\mu&T_{1}&...\\
.&.&.&.&.&...\\
.&.&.&.&.&...\\
\end{array}
\right).
\end{eqnarray}
The matrix elements carry momentum dependence because of the partial Fourier transformation, $T_{1,{\bar 1}} =  2(t\mp\Delta) \cos(\sqrt{3}k_{y}/2)$ and $T_{2,{\bar 2}} = t\pm\Delta$. The appearance of non-zero $T_{\bar 2}$ and $T_{2}$ is the signature of the underlying tripartite lattice.

\begin{figure}
\centering
\includegraphics[width=\columnwidth]{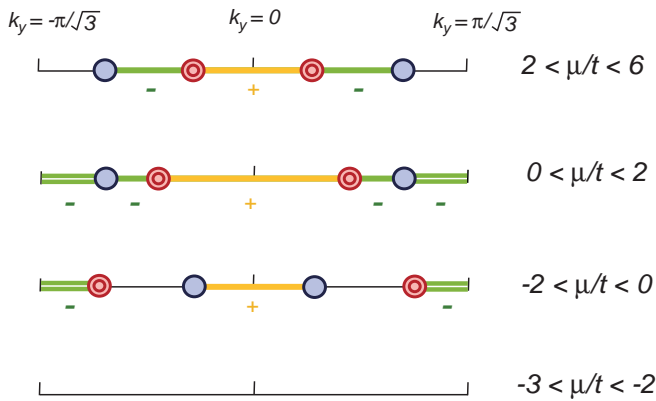}
\caption{\label{fig:EdgeStates} Phase diagram for Andreev edge states. The single circles mark the nodal points without degeneracy while the double circles denote those with two-fold degeneracy. The thick single dashed (yellow) and solid (green) lines indicate the edge state of different particle-hole parities, while the double line denote two independent edge states.}
\end{figure}

It is straightforward to show that zero-energy states exhibit a ``nodal structure'',
\begin{eqnarray}
|\Psi_{-}\rangle = \left(\begin{array}{c}
0\\
\phi_{-}(x)
\end{array} \right),
\qquad
|\Psi_{+}\rangle = \left(\begin{array}{c}
\phi_{+}(x)
\\
0
\end{array} \right).
\end{eqnarray} 
Here $\phi_{-}(x)$ and $\phi_{+}(x)$ belong to the kernel space of the semi-infinite matrix $A$ and $A^{\dag}$ respectively. The edge state can be constructed from the generalized Bloch states, $\phi_{-}(x) = \sum_{i} a_{i} (z_{i})^{x}$, where $z$ satisfies
\begin{eqnarray}
T_{2} z^{2} + T_{1} z 
+ T_{\bar 1} \frac{1}{z}+T_{\bar 2} \frac{1}{z^{2}} = \mu.
\end{eqnarray}
In general, the above equation has four solutions $z_i$. However, the open boundary requires
\begin{eqnarray}
\phi_{-}(-1)=0,\quad  \phi_{-}(0)=0, \quad |\phi_{-}(\infty)|<\infty.
\end{eqnarray}
To satisfy the boundary conditions, we need at least three $|z_{i}| <1$ solutions to construct one edge state $\phi_{-}(x)$. Otherwise, the edge state would be absent. If all of the four solutions satisfy $|z_{i}| <1$, we end up with two edge states. It is clear that we can find $\phi_{+}(x)$ in the similar way. The resultant phase diagram of the AES is shown  in Fig.~\ref{fig:EdgeStates}. The detailed derivation of these solutions will be presented elsewhere\cite{PeregBarnea04}.

\begin{figure}
\includegraphics[width=\columnwidth]{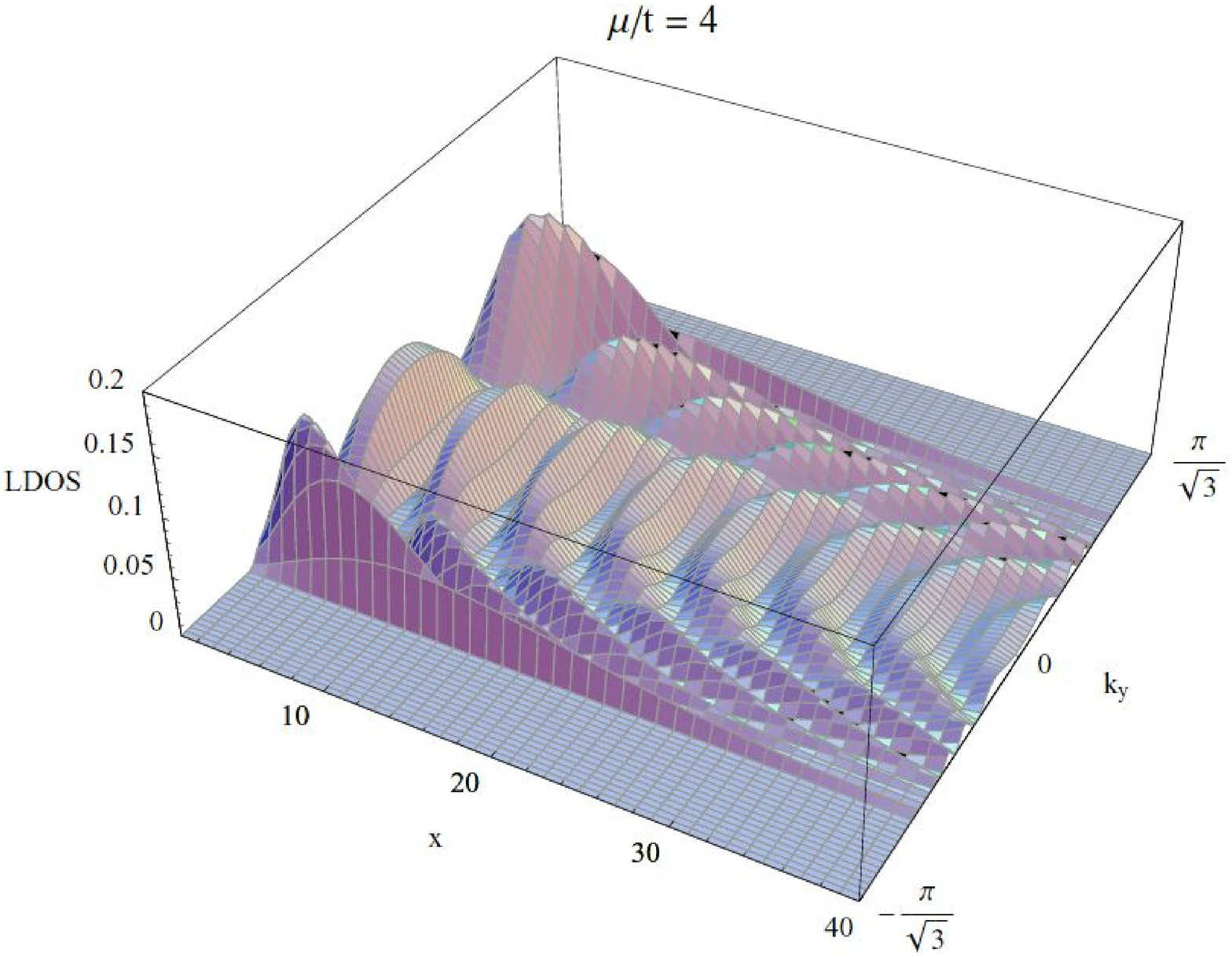}\\
\includegraphics[width=\columnwidth]{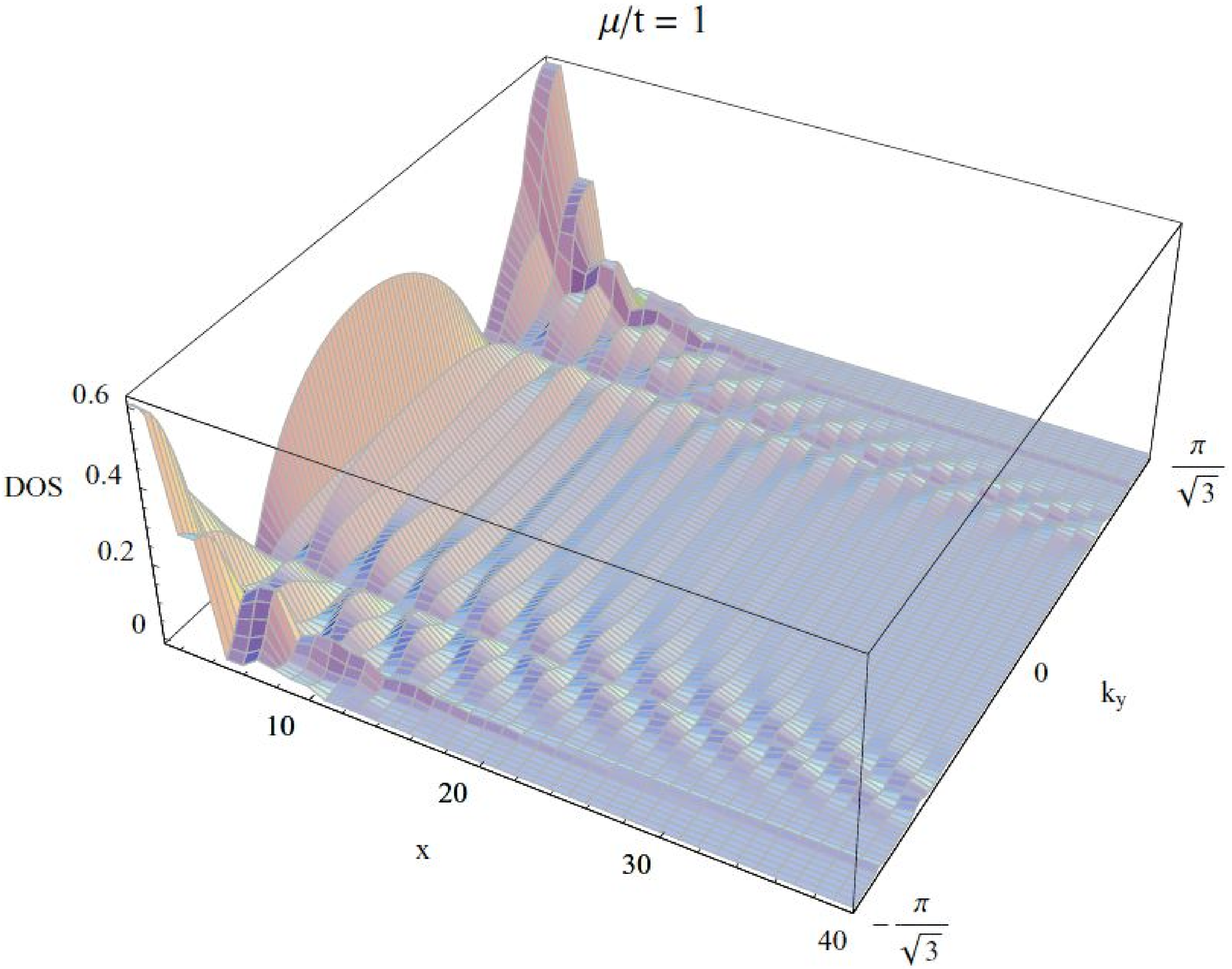}\\
\includegraphics[width=\columnwidth]{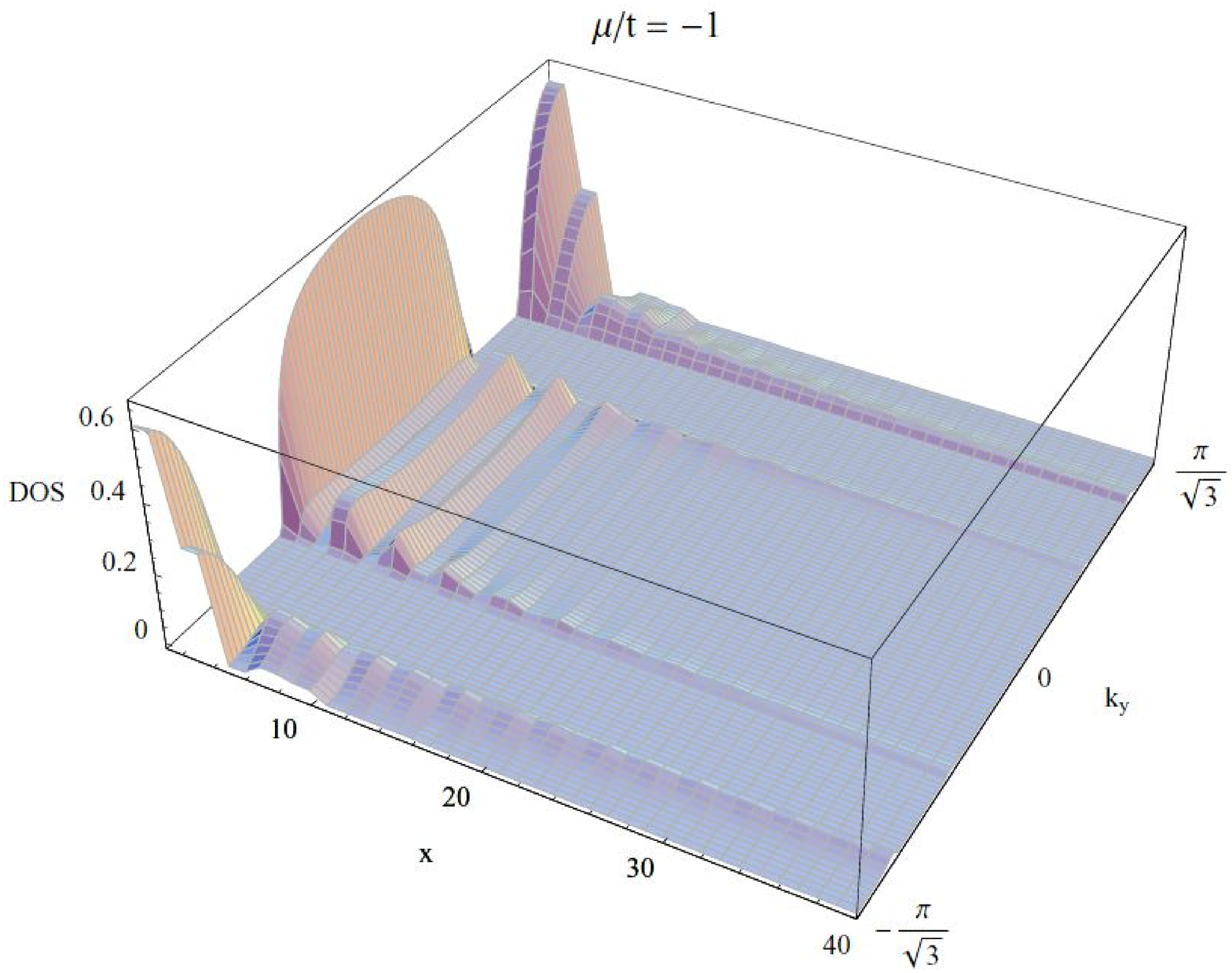}
\caption{\label{fig:LDOS} Local density of states $D(x,k_{y})$ at different chemical potentials. The parameters are chosen to be, $t=1$ and $\Delta=0.4$, in order to visualize the momentum profile. The features are robust against small changes in $\Delta/t$.}
\end{figure}

Before diving into the details of the phase diagram, one notices that an edge state of a certain kind (i.e., $|\Psi_{+}\rangle$ or $|\Psi_{-}\rangle$) {\em always} starts/ends at nodal points. This feature is inaccessible by the conventional Andreev equations in the continuous limit which breaks down near the nodal points. Interestingly, this global structure of the phase diagram is rather general for zero-energy states and can be explained elegantly by the modified Oshikawa's gauge argument\cite{Oshikawa00,Lin04}.

Suppose we wrap up the semi-infinite lattice into tubural conformation and thread a unit flux $\Phi_{0} =2\pi$ through it. After a large gauge transformation to eliminate the flux, the original edge state is mapped into another one with momentum shift $\Delta k_{y} = 2\pi/L$, where $L$ is the circumference of the tube. All edge states between nodal points can be mapped into each other by the flux-gauge transformation because the energy spectrum is gapped. This simple but elegant argument explains the interesting structure in Fig.~\ref{fig:EdgeStates}.

The richness of the phase diagram brings up another puzzle. The existence of the edge state is often understood as the sign change of order parameter at the open boundary, which is robust without detailed dependence on $k_{y}$. Thus, we expect that the phase diagram would be rather simple in the whole Brillouin zone. For simpler systems like the $d$-wave superconductor on the square lattice, this sort of understanding seems to hold rather well. So, what causes the complications on the triangular lattice? The puzzle can be answered by mapping the problem into the effective 1D nearest neighbor tight-binding model.

Any lattice system in arbitrary dimension with open boundary at $x=0$, which is described by a quadratic Hamiltonian can be mapped into a sum of ($k_y$-dependent) one-dimensional models. By choosing an appropriate unit cell the one dimensional chain will contain {\em only} nearest neighbor hopping and in general their Hamiltonian is then given by
\begin{eqnarray}
H = {\bf C}^{\dag}_1 \otimes {\bf R} + {\bf C}_1^{} \otimes {\bf R}^{\dag} + {\bf C}_0 \otimes {\bf 1},
\end{eqnarray}
where ${\bf C}_{1}$ is the hopping matrix connecting nearest-neighbor cells and ${\bf C}_{0}={\bf C}_{0}^{\dag}$ for hopping within the cell. The matrices ${\bf C}_0$ and ${\bf C}_1$ are square matrices with $s$ rows, where $s$ is
the number of effective lattice sites in the unit cell. The semi-infinite matrix $({\bf R})_{x,x'} = \delta_{x+1,x'}$ is the displacement operator on the lattice. 

Again, the edge state can be constructed from Bloch states, $\mybm{\Phi}(x)= \sum_{i} a_{i} \mybm{\phi}_{i} (z_{i})^{x}$, where $z$ satisfies ${\rm det} \left| z {\bf C}_1^{\dag}+\frac{1}{z} {\bf C}_1^{}+{\bf C}_0 \right|=0$. The boundary condition is extremely simple in this representation,
\begin{eqnarray}
{\bf C}_{1} \mybm{\Psi}(0) = \sum_{i} {\bf C}_{1} (a_{i} \mybm{\phi}_{i}) =0.
\end{eqnarray}
Therefore, the number of edge states is exactly the dimension of subspace of the kernel space of ${\bf C}_{1}$, which is spanned by the vectors $a_{i} \mybm{\phi}_{i}$. Consequently, the maximum number of edge states is just the dimension of the kernel space of ${\bf C}_{1}$. On the other hand, if the rank of matrix ${\bf C}_{1}$ is full, it implies no edge. In fact, the reflection symmetry with respect to the open boundary often implies that the rank of 
${\bf C}_{1}$ is full\cite{PeregBarnea04}. For this reason, in some edge topologies, there is never an edge state.

We are now ready to zoom into the details of the phase diagram. To visualize these edge states better, we calculated the local density of states at zero energy, $D(x,k_{y})= \sum_{i} |\Psi_{i}(x)|^{2} \delta(E)$, as shown in Fig.~\ref{fig:LDOS}. For $2<\mu/t <6$, there are three maxima in the LDOS arising from edge states with opposite parities. Note that the edge state merges into the bulk at the nodal points and the weight of the LDOS is suppressed to zero. The (electron) filling factor $x=1.35$ in Na$_{x}$CoO$_{2}$ compound corresponds to the regime $0<\mu/t<2$ where the phase diagram is quite complicated. However, the locations of the LDOS maxima are rather simple, $k_{y} = \pm \pi/\sqrt{3}, 0$. Therefore, we expect to see a sharp peak at $Q_{y} = \pm \pi/\sqrt{3}a$ in the STM data after Fourier analysis. For $-2<\mu/t<0$, the structure of LDOS is pretty much the same except for an additional regime where the LDOS vanishes. 
Again, it would predict the same sharp peak at $Q_{y} = \pm \pi/\sqrt{3}a$ in the STM data. Finally, for $-3<\mu/t<-2$, there is no edge state.

The approach developed here can be directly applied to other pairing symmetries such as $p$- or $d$-wave. Since these pairing symmetries do not match the hexagonal symmetry, the result sensitively depends on 
the orientations of the gap functions relative to the underlying lattice. 
As in the high $T_{c}$ materials, the correlation length in Na$_{x}$CoO$_{2}$ is small\cite{Chou04}. It is then reasonable to assume that the orientation of the gap function is pinned to the lattice, i.e. either the maximal gap line or the nodal line would lay on a lattice bond. This gives rise to eight possibilities (shown in Fig.~\ref{fig:cuts}) of $p_{x}$, $p_{y}$, $d_{xy}$ and $d_{x^{2}\mbox{-}y^{2}}$ symmetries at zigzag and flat edges. Since the calculations are rather straightforward, the results are summarized in Table \ref{PairingSymmetry} with detailed phase diagrams deferred for future publication.

Our approach presented in this Letter ignores the disorder and correlation effects. While it was demonstrated in the literature that the weak impurities and dislocations do not destroy the AES, correlation effects can be more dramatic. In fact, it is rather interesting to explore some exotic phenomena 
such as spin-charge separation in these effective one-dimensional system, which may be rather different from the true ladder-like materials. Another extension of our work would be to include the phase fluctuations of the order 
parameter and formulate the effective theory in terms of gauge fields. 
In summary, we found that the AES reveals rich and interesting structures which can serve as a good indicator for the pairing symmetry. The phase diagram can be understood by the more general mapping/argument, suggesting that the phenomena presented here are robust with respect to minor details. Further studies to include the correlation effects would shed light on interesting bulk physics by probing the edge.

We acknowledge Leon Balents, Matthew Fisher, Marcel Franz, Chung-Yu Mou, Gil Rafael and Sungkit Yip for useful discussions. HHL appreciates financial supports from National Science Council in Taiwan through Ta-You Wu Fellow and grants NSC-91-2120-M-007-001 and NSC-92-2112-M-007-039. TPB wishes to acknowledge NSERC for financial support. The hospitality of KITP, where most of the work was done, is greatly appreciated.

\end{document}